\documentclass[11pt,preprint]{aastex}

\def\t0{\theta_{\circ}}

\def\be{\begin{equation}}
\def\en{\end{equation}}

\def\msun{M_{\sun}}
\def\rsun{R_{\sun}}
\def\lsun{L_{\sun}}
\def\msunyr{M_{\sun} \, yr^{-1}}

\def\mdot{\dot{M}}

\def\curf{{\cal F}}
\def\ecs{erg\;  s^{-1}\;cm^{-2}}

\def\h2{H$_2$}

\begin{document}

\title
{Using FUV to IR Variability to Probe the Star--Disk Connection in the Transitional Disk of GM Aur}

\author{Laura Ingleby\altaffilmark{1}, Catherine Espaillat\altaffilmark{1}, Nuria Calvet\altaffilmark{2}, Michael Sitko\altaffilmark{3}, Ray Russell\altaffilmark{4}, Elizabeth Champney\altaffilmark{3}}

\altaffiltext{1}{Department of Astronomy, Boston University, 725 Commonwealth Avenue, Boston, MA 02215, USA; lingleby@bu.edu}
\altaffiltext{2}{Department of Astronomy, University of Michigan, 830 Dennison Building, 500 Church Street, Ann Arbor, MI 48109, USA}
\altaffiltext{3}{Department of Physics, University of Cincinnati, Cincinnati OH 45221, USA }
\altaffiltext{4}{The Aerospace Corporation, Los Angeles, CA 90009, USA}

\begin{abstract}
We analyze 3 epochs of ultraviolet (UV), optical and near-infrared
(NIR) observations of the Taurus transitional disk GM Aur using
the $\emph{Hubble Space Telescope}$ Imaging Spectrograph (STIS) and
the $\emph{Infrared Telescope Facility}$ SpeX spectrograph.
Observations were separated by one week and 3 months in order to study
variability over multiple timescales.  We calculate accretion rates
for each epoch of observations using the STIS spectra and find that
those separated by one week had similar accretion rates ($\sim1\times10^{-8}$ $\msunyr$) while the
epoch obtained 3 months later had a substantially lower accretion rate ($\sim4\times10^{-9}$ $\msunyr$).
We find that the decline in accretion rate is caused by lower
densities of material in the accretion flows, as opposed to a lower
surface coverage of the accretion columns.  During the low accretion
rate epoch we also observe lower fluxes at both far UV (FUV) and IR wavelengths,
which trace molecular gas and dust in the disk, respectively.  We
find that this can be explained by a lower dust and gas mass in the inner disk.
We attribute the observed variability to inhomogeneities in the inner disk, near the corotation radius, where gas and dust may co-exist near the footprints of the magnetospheric flows.  These FUV--NIR data offer a new perspective on the structure
of the inner disk, the stellar magnetosphere, and their interaction.

\end{abstract}


\section{ Introduction}
\label{intro}
The early picture of static, axisymmetric circumstellar disks
irradiated by a constant luminosity T Tauri star has evolved to
include disk warps, fluctuating accretion of material in the disk and
ever changing emission properties of the central star, among other
dynamic processes.  Each of these manifest themselves in T Tauri light
curves as variability.  Warps block light from the star, causing
optical dips \citep{bouvier99,alencar10,morales11,cody14} while
variable accretion may be seen as bursts in optical and ultraviolet
light curves \citep{herbst94,stauffer14} or changing widths and
intensities of emission lines produced by accretion
\citep{costigan12,chou13,dupree14}.  Activity cycles in the star may
cause long timescale X-ray variability \citep{micela03}, while short,
strong X-ray bursts are produced by solar-like flares \citep{wolk05}.
Depending on the cause, variability is seen as periodic or stochastic,
lasting for minutes or slowly varying over a period of years.  With
all these characteristics of T Tauri light curves, often with multiple
phenomena occurring simultaneously, explaining variability is
complicated, especially if the time sampling is low or the wavelength
coverage is limited.

In addition, the radial structure of dust in the disk is not always
continuous.  Along with full disks, those with gaps or holes in the
dust (pre-transitional or transitional, respectively) are known to
exist \citep{strom89,skrutskie90,calvet02,espaillat07,andrews11} all
thought to eventually end up as debris disks with little gas and
processed dust \citep{zuckerman01,chen06,matthews14}.  Planet
formation, photoevaporation of the disk by the central star or viscous
evolution, and settling have all been invoked to explain these
structures observed in the dust \citep{espaillat14,zhu11,owen12}.
Gaps and holes in the disk can alter interpretations of T Tauri
variability, as processes relevant to full disks appear differently in
transitional disks.  For example, a warp in the disk wall at the
sublimation temperature may not be a contributor to the light curves 
of transitional disks, which have walls at much larger radii.  Therefore,
varied dust distributions add an additional complication to the
interpretation of T Tauri variability.

Here, we analyze variability in the transitional disk GM Aur over 3
epochs of observation with full wavelength coverage from the far-ultraviolet (FUV) to
near-infrared (NIR), all obtained within one day.  While photometric variability
studies focusing on one or multiple wavelength regimes are significant
in number, there are few, if any, variability studies with such
extensive spectral coverage.  GM Aur is a K5 source in the Taurus
molecular cloud, initially suspected to be a transitional disk
\citep{marsh92,koerner93,chiang99} and later established as having an
inner clearing of optically thick dust, though with a small amount of
submicron-sized optically thin dust remaining, using \emph{Spitzer}
Infrared Spectrograph (IRS) observations
\citep{calvet05b,espaillat10}.  Sub-millimeter observations confirmed
that GM Aur has an inner hole in the dust, with a radius of $\sim$ 20
AU \citep{hughes09,andrews11}.  \citet{rice03} showed that
transitional disk IRS spectra may be produced by clearing of disk
material by planets over a range of disk radii.  While
photoevaporation models can also produce inner disk holes, the size of
the cavity in GM Aur and the presence of remaining gas and small dust
are better understood in a planet formation scenario \citep{owen10}.
In addition to existing long wavelength observations of GM Aur, far
and near ultraviolet (FUV and NUV, respectively) spectra have provided
glimpses of the gas in the disk of GM Aur \citep{ardila13} and the
emission produced in an accretion shock at the stellar surface
\citep{ingleby13}.

In \S \ref{obs}, we discuss the UV, optical and near- IR
observations used in our analysis.  In \S \ref{analysis} we outline
our modeling procedures for each epoch of GM Aur data, including
accretion shock modeling, comparison of IR fluxes to models of the
emission from dust in the disk, and a discussion of the molecular
contribution to the FUV.  Finally, in \S \ref{discussion} we
propose a scenario to explain the variability of these spectra.

\section{Observations}
\label{obs}
\subsection{STIS}
\label{stis}
\begin{figure}[htp]
\plotone{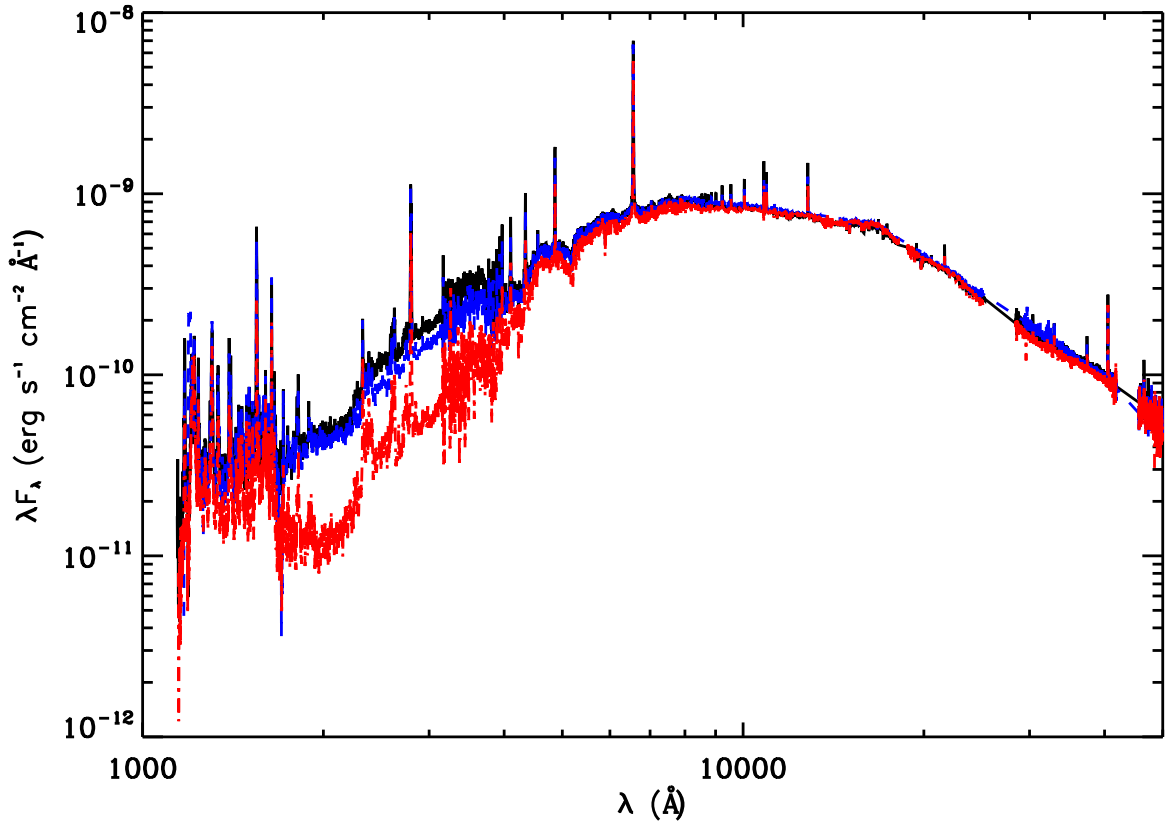}
\caption{UV, optical and IR spectra of GM Aur.  Three epochs of
observations are shown as solid black (GM\_1), dashed blue (GM\_2) and
dot-dashed red (GM\_3) lines.  Gaps in the near- IR SpeX data are
regions where data is not shown due to poor telluric subtraction.}
\label{all}
\end{figure}

GM Aur was observed during 3 epochs, referred to as GM\_1, GM\_2 and
GM\_3 in the following analysis, between September 2011 and January
2012 in GO program 11608 (PI: N. Calvet) with the Space Telescope
Imaging Spectrograph (STIS).  The observations were separated by 1
week (GM\_1 to GM\_2) and 3 months (GM\_1 to GM\_3) in order to study
variability on short and long timescales.  Spectra were obtained from
the FUV to optical, approximately 1100 {\AA} to 1 $\mu$m, utilizing a
combination of STIS gratings.  The UV gratings, G140L and G230L used
the MAMA detector and provided spectra between 1150 and 3180 {\AA}
with resolution, $R$, ranging between 500 and 1,440.  G430L and G750L
used the CCD detector and covered 2900 to 10,270 {\AA} with
$R\sim530-1,040$.  We used data products produced through the STScI
\emph{calstis} reduction pipeline for our analysis.  A log of all STIS
observations is provided in Table \ref{tabobs} and the spectra are
plotted in Figure \ref{all}, combined with IR data obtained from SpeX,
discussed in \S \ref{spex}.

\subsection{SpeX}
\label{spex}
GM Aur was observed using the SpeX spectrograph \citep{rayner03} on
NASA's Infrared Telescope Facility (IRTF). The SpeX observations were
obtained using the cross-dispersed (hereafter XD) echelle gratings in
both short-wavelength mode (SXD) covering 0.8-2.4 \micron{} with a resolution of $\sim$2000 and
long-wavelength mode (LXD) covering 2.3-5.4 \micron{} and resolution $\sim$2500.  The LXD data
were obtained on 5 nights (2011 September 11, 18, and 29 (UT) and 2012
January 2 and 6 (UT). On three of these nights (November 11 \& 18 and
January 6) SXD data were also obtained. For all XD observations, a
0.8$''$ wide slit was used. The spectra were corrected for telluric
extinction, and flux calibrated against the A0V star HD 25152 using
the Spextool software \citep{vacca03,cushing04} running under IDL. On
nights when both SXD and LXD spectra were obtained, the LXD spectrum
was merged to the SXD, by re-scaling in the LXD spectra in the
wavelength region where they overlapped.

Due to the light loss introduced by the 0.8$''$ slit used to obtain
the XD spectra, changes in telescope tracking and seeing between the
observations of GM Aur and a calibration star may result in merged XD
spectrum with a net zero-point shift compared to their true absolute
flux values. To correct for any zero-point shift that results from the
0.8$''$ echelle observations, we also observed GM Aur on 4 of the
nights (excluding only 2011 September 29, due to poor sky conditions)
with the low dispersion prism in SpeX using a 3.0$''$ wide slit. On
nights where the seeing is 1.0$''$ or better, this technique yields
absolute fluxes that agree with aperture photometry to within 5\% or
better.  The resulting scaling factors derived for the XD spectra were
10\% or less on all 4 nights, and the scalings were applied to each
spectrum prior to analysis.  The three epochs closest in time to the
STIS spectra were included in our analysis and details of the SpeX
observations are listed in Table \ref{tabobs}.

\section{Analysis and Results}
\label{analysis}

\subsection{Accretion rates from STIS spectra}
\label{accretion}
It is generally accepted that the NUV and blue optical excess emission in
classical T Tauri stars (CTTS) is produced in the accretion shock at
the stellar surface, a result of magnetospheric accretion
\citep{uchida85,koenigl91,shu94}.  In this scenario the circumstellar
disk is truncated by strong stellar magnetic fields at a few stellar
radii, inside of which material is channeled along the field lines to
the star.  \citet{calvet98} described the emission from an accreting
column of material which forms a shock at the stellar surface.  The
shock emits soft X-rays towards the star, heating the photosphere
below, and along the column of accreting material where it is
reprocessed producing a spectrum peaked in the NUV.  Models which
assume a single temperature and density slab for the accretion shock
have also been successful in reproducing the shock emission
\citep{herczeg08, rigliaco11, manara14}.  As a second order
approximation to the accretion shock model of \citet{calvet98},
\citet{ingleby13} allowed columns with a variety of energy fluxes,
$\curf$, and surface coverage or filling factors, $f$, to co-exist.
The presence of multiple accretion columns was motivated by models of
the magnetospheric geometry which include higher order moments of the
stellar magnetic field producing a complex accretion pattern at the
stellar surface \citep{donati08,long08,gregory11}.

We fit the STIS NUV and optical spectra with the multi-column
accretion shock model described in \citet{ingleby13}.  STIS spectra
were first de-reddened using $A_V=0.8$ from \citet{espaillat11} and
the extinction law towards HD 29647 \citep{whittet04}.  We account for
the emission from a T Tauri chromosphere and photosphere by using a
non-accreting T Tauri star or weak T Tauri star (WTTS) template, specifically
a STIS spectrum of the WTTS RECX 1 from \citet{ingleby11b}.  RECX 1 is
in the nearby $\eta$ Chamaeleon star forming region and has the same
spectral type as GM Aur (K5).  

To scale the spectrum of the WTTS RECX 1 to that of GM Aur we use the
veiling value at the V band.
Veiling refers to the degree by which photospheric absorption lines
appear shallower due to the addition of a continuum excess
\citep{hartigan91}, which we
assume is accretion shock emission, or
$F_{V,WTTS}=F_{V,CTTS}/(1+r_V)$, where $F_{V,WTTS}$ and $F_{V,CTTS}$
are continuum fluxes for the WTTS and CTTS, respectively, and $r_V$ is
the amount of veiling, defined as $r_V=F_{V,Veil}/F_{V,WTTS}$, where
$F_{V,Veil}$ is the flux of the veiling continuum.
Since we lack high resolution observations, we use
the value $r_V=0.2$, from \citet{edwards06}.  Veiling is variable and depends on the amount of excess emission produced in the shock, therefore assuming the value of \citet{edwards06} is a source of error.  However, we expect this error to be small given that the low accretion rate of GM Aur produces a small amount of veiling.  The optical spectrum of GM Aur is dominated by the stellar emission, with the shock emission becoming important in the UV.

As in \citet{ingleby13},
each accretion column is characterized by an energy flux, $\curf$, 
and a filling factor, $f$. 
The energy flux, $\curf=1/2\rho v_s^3$, is a
measure of the density of material in the accretion column, $\rho$,
assuming that the magnetospheric truncation radius is not changing and
therefore the infall velocity, $v_s$ is constant.  
Higher $\curf$
columns deposit more energy, producing hotter shock emission and
shifting the shock spectrum towards shorter wavelengths \citep{ingleby13}.
The filling factor $f$ gives the fraction of the stellar surface covered
by the column.
The stellar mass and radius are additional 
input parameters to the models.
We estimate stellar properties using a synthetic $J$ band
 magnitude found by integrating the 2MASS $J$ band transmission curve
 over the de-reddened SpeX observations.  The synthetic $J$ magnitude
 combined with the conversion to bolometric magnitude of
 \citet{kenyon95} provides the stellar luminosity,
 $L_{\ast}=0.9\;\lsun$.  We assume a distance to GM Aur of 140 pc,
 and a temperature of 4350 K
 \citep{kenyon95} 
 to
 get $R_{\ast}=1.7\;\rsun$.  Finally we estimate a mass 
$M_{\ast}=1.1\;\msun$
using the isochrones of \citet{siess00}. 

\begin{figure}[htp]
\plotone{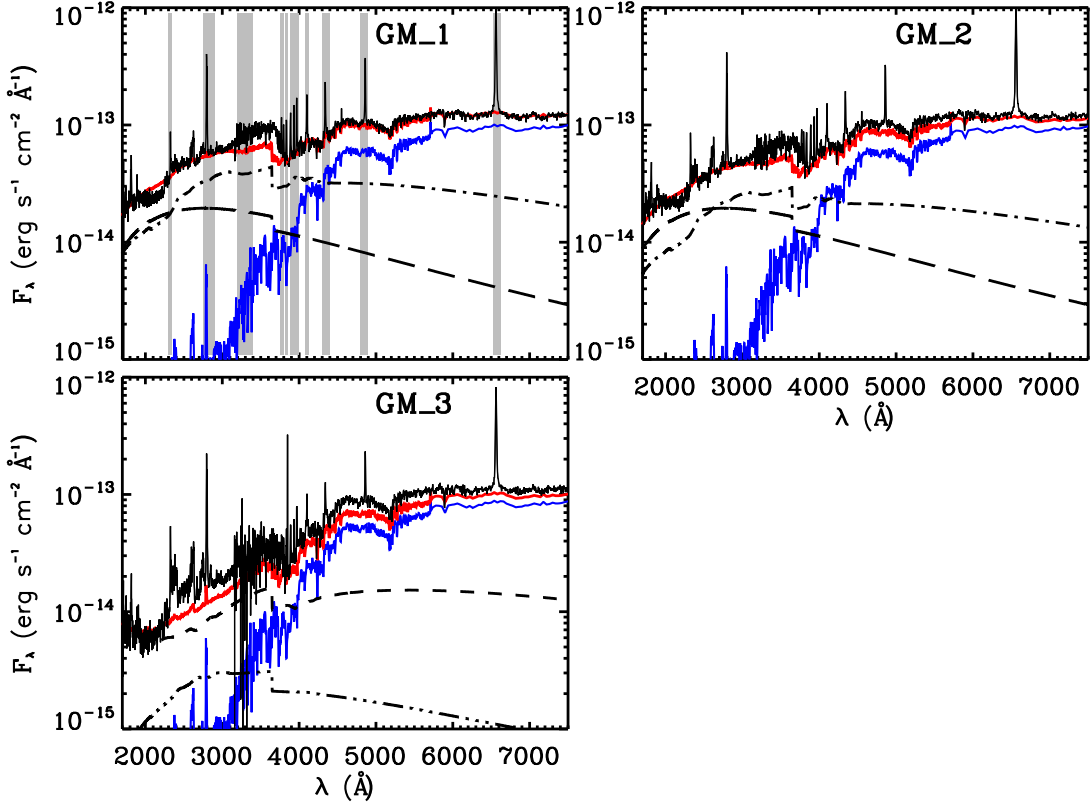}
\caption{Accretion shock model fits to 3 epochs of GM Aur STIS
spectra.  In each panel the black line is the de-reddened STIS
spectrum, while the blue line is a STIS WTTS template, here RECX 1.
The broken lines represent accretion columns characterized by
different values of $\curf$; log $\curf$=10, 10.5 (short dashed), 11
(dot-dashed), 11.5 (dash-triple dotted), and 12 (long dashed) erg
s$^{-1}$ cm$^{-2}$.  
No columns of log $\curf$=10 erg s$^{-1}$
cm$^{-2}$ were needed for the fits.  The red line is the best fit to
the data of the WTTS template plus the summed accretion column emission.  The gray regions in the first panel represent spectral regions excluded when calculating the model fit to the data. }
\label{shocks}
\end{figure}

We calculated models of the emission from accretion columns
with log $\curf$=10, 10.5, 11, 11.5 and 12 erg s$^{-1}$ cm$^{-2}$, corresponding to heated photosphere temperatures between 4600 and 10,700 K.  As
$\curf$ increases, the shock spectrum gets bluer, therefore, a
combination of columns provides the best fit to red and blue
wavelengths.  
Following \citet{ingleby13},
we found the best model of the STIS spectra by
calculating the reduced chi squared fit ($\chi^2_{red}$) of combinations of accretion columns,
each scaled by $f$, left as a free parameter.  The accretion shock models do not attempt to reproduce line emission, therefore, the $\chi^2_{red}$ fit omitted spectral regions with strong emission lines.  The assessment of the fit also excluded the region between 3200 and 3400 {\AA}, where there is significant noise due to edges of the NUV and optical gratings.   To produce the final
fit, we added the emission from each contributing accretion column
to the emission from the WTTS template, RECX 1.  Figure \ref{shocks}
shows the best fits to each epoch of STIS observations.

The filling factor of the contributing columns at each
epoch is given in Table \ref{tabacc}. 
We list a characteristic value for the energy flux, $\curf_w$,
which is an average of the energy flux
of all columns, weighted by the filling factor of each
column.  The listed total filling factor, $f_{total}$, is the sum of the $f$ values for the
contributing columns, providing the total percentage of the surface
covered by accretion columns.  

The mass accretion rate
was found by 
by adding the contributions of the columns \citep{ingleby13}
 \be
\mdot=\frac{8\pi R_{\ast}^2}{v_s^2} \sum_i \curf_i f_i =
\frac{8\pi R_{\ast}^2}{v_s^2} \curf_w f_{total} 
 \en
We assume that $v_s$ is 0.89 of the free-fall velocity, which
for the stellar parameters is
494 ${\rm km \, s^{-1}}$, corresponding to infall from
$\sim$ 5 stellar radii.
Values for $\mdot$ in the three epochs are given in 
Table \ref{tabacc}.  The error in our estimates of the accretion rate is dominated by uncertainties in $A_V$.  Existing estimates of extinction for GM Aur range from 0.1 \citep{kenyon95} to 0.8 \citep{espaillat11} with a value of 0.3 derived from long wavelength optical spectra \citep{herczeg14}.  We adopted the value listed in \citet{espaillat11} in order to directly compare those results to our analysis.  The range of measured $A_V$ values introduces a factor of 2 error in our $\mdot$ estimate.

\begin{figure}[htp]
\plotone{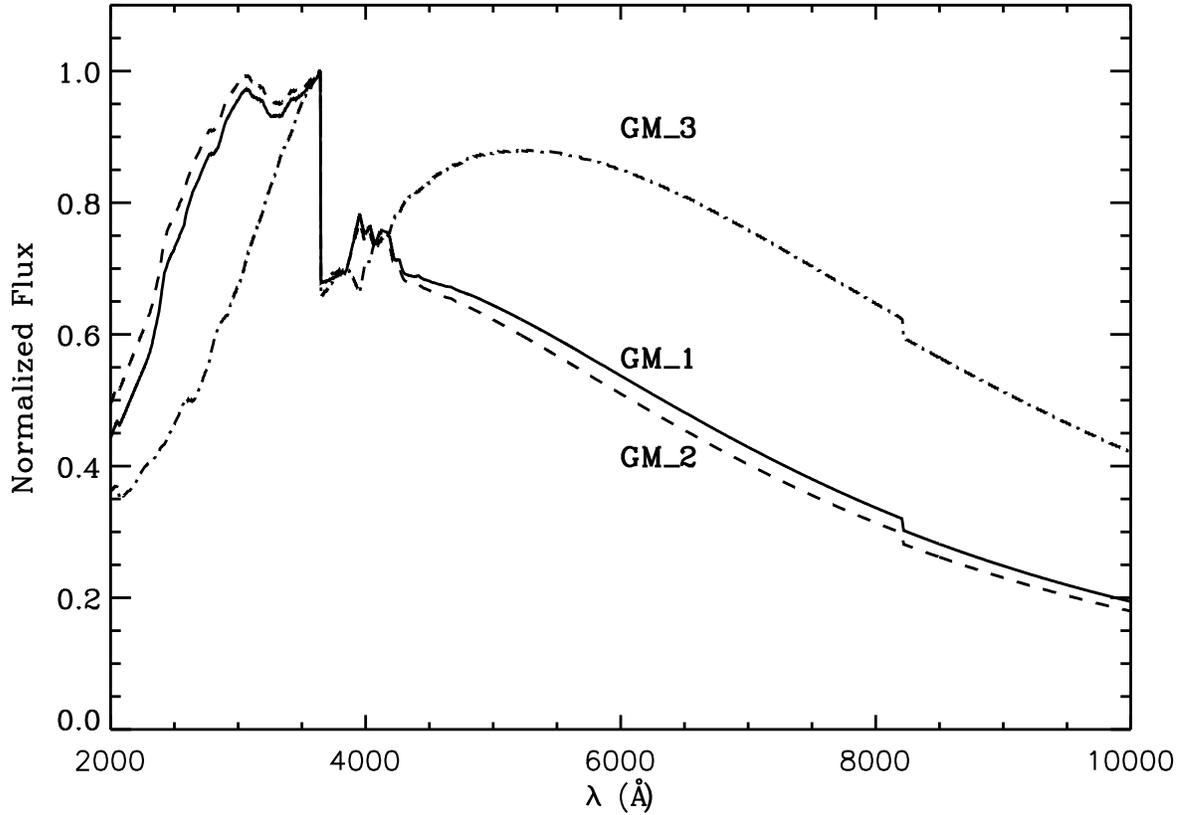}
\caption{Total shock models used to fit the STIS spectra.  The
contribution from each accretion column in the fit is summed for the
final accretion shock models.  Fluxes are scaled for direct comparison
of the shape of the shock emission.  The emission from the accretion
shock is characterized by higher density columns for GM\_1 (solid) and
GM\_2 (dashed), or higher $\curf$.  GM\_3 (dash-dotted) has the lowest
density of accreting material and therefore the coolest shock
spectrum.}
\label{curf}
\end{figure} 

\subsection{FUV to IR variability}
Here, we analyze the wavelength dependent variability in the STIS and
SpeX data and discuss some of the common contributors to the emission
in each spectral region.  GM\_1 and GM\_2, separated by one week, have
very similar spectra and in \S \ref{accretion} we showed that
these two epochs had minimal differences in the accretion properties.
However, GM\_3, which was observed 3 months after GM\_1, is clearly
dimmer, though the magnitude of the dimming across the spectrum is not
uniform (see Figure \ref{all}).  Namely, the biggest decrease in
fluxes occurs in the NUV, with accompanying yet smaller dips in FUV,
optical and IR fluxes.  In the following analysis we focus on the
differences between GM\_1 and GM\_3.

\begin{figure}[htp]
\plotone{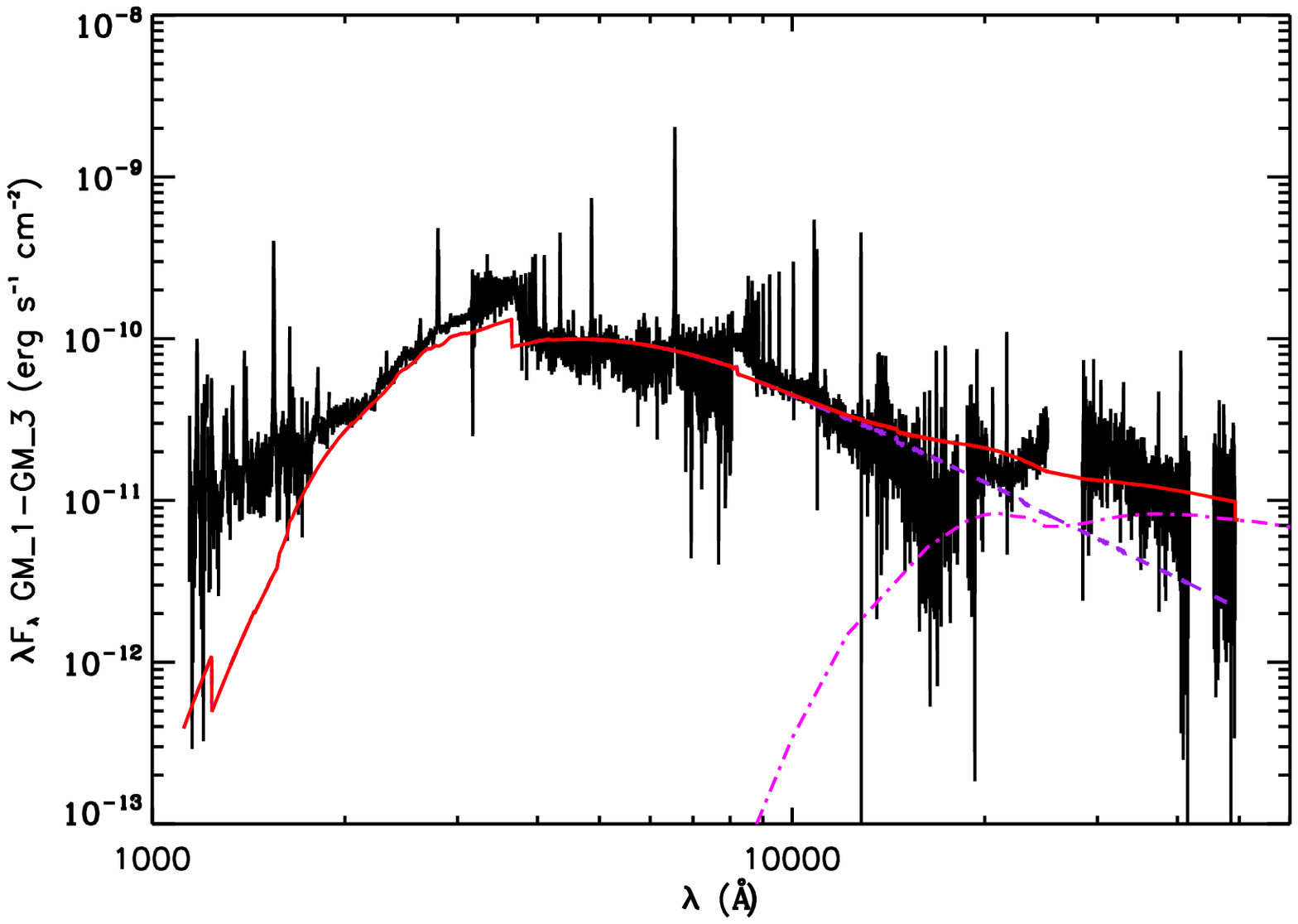}
\caption{Excess in GM\_1 over GM\_3.  The spectrum of GM\_3 was
subtracted from GM\_1 to isolate the excess emission observed in the
first two epochs (solid black line).  The excess is fit in the NUV and
optical with accretion shock emission characterized by
$\curf=3\times10^{11}$ erg s$^{-1}$ cm$^{-2}$ and $f=0.006$ (dashed
purple line).  The optically thin dust emission is also shown to
contribute to the IR emission (dash dotted, magenta line).  The solid
red line is the total of the accretion shock emission and the
optically thin dust emission.  Here we are not attempting to explain
the FUV excess.}
\label{diff}
\end{figure} 

\subsubsection{NUV and optical excess}
\label{nuvopt}

As shown in Table \ref{tabacc},  
while the total filling factors for all three
epochs are approximately equal, 
the energy flux of the columns is higher in
GM\_1 and
GM\_2 than in GM\_3. As a result, 
the columns in 
GM\_3 emit less in the NUV, significantly
lowering the total flux in this
wavelength range. This can best be seen in   
Figure \ref{curf}, which compares the normalized
total
flux due to the accretion columns in the
three epochs.

Figure \ref{diff} shows the difference in flux between GM\_1 and
GM\_3, with excesses observed in GM\_1 at all wavelengths.  
We fit the NUV and
optical portion of this difference spectrum with the emission from an
accretion column.  We tried fits with columns characterized by log
$\curf=10-12$ and found that the best fit to the excess comes from an
accretion column with $\curf=3\times10^{11}$ $\ecs$ and $f$ of 0.006.
The accretion rate of a column with these parameters is
5.1$\times10^{-9}\;\msunyr$.  This $\mdot$ is roughly consistent with
the difference in $\mdot$ between GM\_1 and GM\_3 listed in Table
\ref{tabacc}.  

To summarize, we find that the difference between the accretion
properties of GM\_1 and GM\_3 is in the energy flux, or density, of
the accretion columns rather than the total surface coverage.  
The decrease of $\mdot$ in GM\_3 is mainly due to the decreased density
in the high density accretion columns.

\subsubsection{IR excess}
\label{ir} 

GM~Aur is known to be variable in the infrared.  Using
{\it Spitzer} IRS spectra, \citet{espaillat11} found a change in
the strength of the 10~{\micron} silicate emission and in infrared
continuum emission from 5 to 10~{\micron}. They interpreted this as due
to changes in the amount of optically thin dust within the disk hole.
\citet{espaillat11} found that the inner disk hole
contained 2$\times$10$^{-12}$ M$_{\sun}$ of optically thin dust and they could
reproduce the observed variability by decreasing the amount of dust in
the optically thin region by a factor of two.

In Figure \ref{diff} we detect an excess in the IR in GM\_1 as
compared to GM\_3.  This excess is redder than the emission expected
from accretion
and therefore not associated with
shock emission \citep{fischer11,mcclure13a}.  
We calculated models of
optically thin dust emission to fit the SpeX data for GM\_1 and GM\_3,
following the procedure of \citet{espaillat10,espaillat11}
to reproduce the excess. The dust is heated by stellar radiation 
and we change the mass in dust to fit the spectra.
The silicate dust composition used in the optically thin
region is the same as in \citet{espaillat11}, which is based on the
best fit to the 10~{\micron} silicate feature. We use ISM-sized
(0.005--0.25~{\micron}) dust composed of 32$\%$ organics, 12$\%$
troilite, and 56$\%$ silicates. The silicates are 90$\%$ amorphous and
10$\%$ crystalline forsterite. 
We find that to fit the GM\_1 spectra
we need 1.9$\times$10$^{-12}$ M$_{\sun}$ of dust in the
optically thin region,
while we need 
1.4$\times$10$^{-12}$ M$_{\sun}$ for GM\_3.  Figure
\ref{diff} compares the GM\_1-GM\_3 excess spectrum to the emission
from the difference between the two optically thin dust models used to
fit the SpeX data.

\subsubsection{FUV excess}
\label{fuv}
Figure
\ref{diff} shows that
the observed FUV excess in GM\_1 compared to
GM\_3  
cannot be explained by
emission from more energetic accretion columns alone.
However, 
it is well known that
in addition to shock emission, 
molecular
gas emission contributes to the FUV spectrum, primarily \h2 with a
contribution from CO
\citep{herczeg02,herczeg04,bergin04,ingleby09,france11a,france11b}.
This molecular gas was expected to be within a few AU of the central
star, as \h2 was spatially unresolved by \citet{herczeg02} for TW Hya,
and indeed
\citet{ingleby11b} showed that spectrally resolved \h2 lines originated
between the magnetospheric truncation radius, out to $\sim$3 AU for
the case of RECX 11 in $\eta$ Chameleon.

\h2 in the inner disk is excited via one of two mechanisms.  The first
is excitation by Ly$\alpha$ photons populating the upper levels of \h2
which then creates a fluorescent spectrum as it de-excites.  While
Ly$\alpha$ fluorescent lines were identified by \citet{herczeg02} in
high resolution STIS spectra, the features are detectable at low
resolution and are found in the GM Aur FUV spectra for all epochs.
The second mechanism is collisional \h2 excitation by free hot
electrons.  Metals present in the inner disk are ionized by X-rays
from the star and these electrons ionize the abundant H and He
producing a large supply of hot electrons.  Electrons collisionally
excite \h2 and therefore different paths for de-excitation are allowed
with one path producing continuum emission as the \h2 is dissociated.
Therefore the two excitation methods produce unique and
distinguishable \h2 emission spectra \citep{abgrall97}.

\begin{figure}[htp]
\plotone{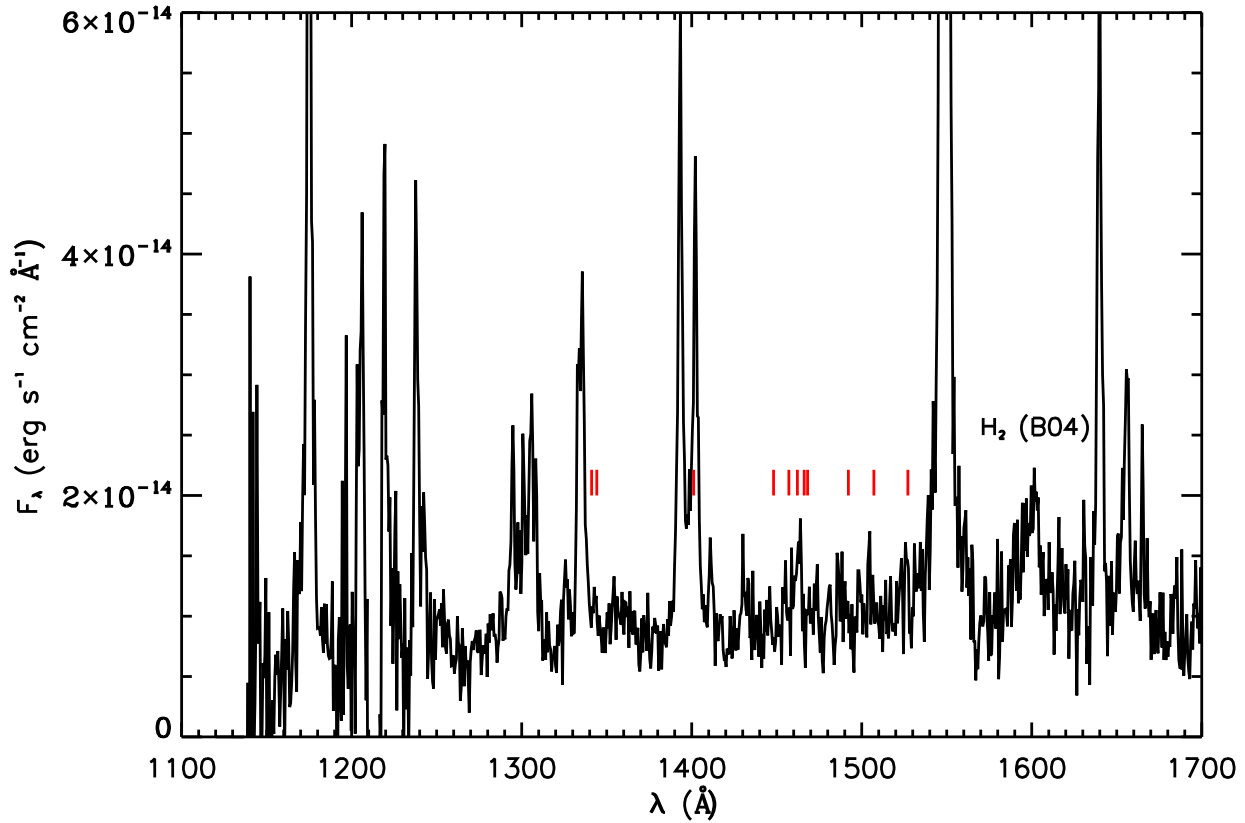}
\caption{FUV excess in GM\_1 - GM\_3 after shock subtraction.  The
black line shows FUV difference spectrum minus the emission from the
accretion column.  A continuum feature discussed in \citet[][B04]{bergin04}
 is identified near 1600 {\AA} and produced by electron impact
excitation of \h2.  The wavelengths of several lines produced by
Ly$\alpha$ fluorescence are marked by red dashes.}
\label{h2}3
\end{figure} 

In Figure \ref{h2} we show the excess FUV emission after the
subtraction of the accretion shock continuum emission (black minus red in Figure
\ref{diff}).  The strong FUV atomic lines remain 
after the removal of the shock, since 
the accretion shock models do not attempt to produce line emission;
however lines such as the Si IV doublet near 1400 {\AA}, C IV at 1548
and 1551 {\AA} and He II $\lambda$1640 {\AA} are known to correlate
with the mass accretion rate \citep{johnskrull00,calvet04,ardila13}.
In addition to the strong hot lines, we see a remaining continuum
emission which cannot be explained by the accretion shock.  The
broad continuum feature at 1600 {\AA} was identified as
collisionally excited \h2 emission from the disk of CTTS in
\citet{bergin04}. \citet{ingleby09,ingleby12} showed that 
this feature is
always observed in CTTS but absent in WTTS, when the disk has been
dispersed.  We observe this continuum feature in the FUV difference
spectrum, revealing an excess of collisionally excited \h2 emission
coming from the disk in GM\_1.
Assuming that the difference reflects an increase in \h2 density,
we estimate
the necessary
density difference assuming that the flux in the 1600 {\AA} \h2
feature ($F_{1600}$) is proportional to the \h2 surface density
($\Sigma_{H_2}$), or $F_{1600}\propto \Sigma_{H_2}^2$
\citep{ingleby09}.  Using this relation, we find that the surface
density of \h2 decreased from GM\_1 to GM\_3 by a factor of 1.4.  

The difference spectrum does not support an excess in
the Ly$\alpha$ excited \h2 emission.  Fluorescent lines easily
detected in the original spectra do not appear in Figure \ref{h2},
indicating that there is little change in this type of \h2 emission
when the accretion rate drops.  In addition to the observations discussed here, GM Aur was
observed with STIS in April, 2003 in GO Program 9374 (PI: E. Bergin)
and during August, 2010 in GO Program 11616 (PI: G. Herczeg).  The
2010 spectra were analyzed and accretion rates determined in
\citet{ingleby13} while we calculate $\mdot$ for the 2003 data using
the method described in \S \ref{accretion}.  Accretion rates in
2003 and 2010 were $9.1\times10^{-9}$ and $9.6\times10^{-9}$
$\msunyr$, respectively.  We compared the fluxes in the strong Ly$\alpha$ fluoresced \h2 lines of all 5 STIS epochs and found that there is no evidence for a correlation with $\mdot$, based on a Pearson r coefficient between 0 and 0.5 for most lines.  While Ly$\alpha$ is heavily absorbed
by interstellar hydrogen and cannot be observed directly, these
results suggest Ly$\alpha$ is not changing between GM\_1 and GM\_3,
contrary to predictions that Ly$\alpha$ is formed in the accretion
shock.  If instead Ly$\alpha$ is formed in the broad accretion flows,
then perhaps the magnetospheric truncation radius is larger
when $\mdot$ drops, increasing the Ly$\alpha$ emitting region.  While
interesting, much more substantial work is needed to interpret
Ly$\alpha$ in CTTS, for example with efforts to reconstruct the
Ly$\alpha$ line profile by observing \h2 fluorescence
\citep{herczeg02,schindhelm12}.

\section{Discussion}
\label{discussion}

In the previous sections, we presented observations which
displayed variability in the FUV--NIR emission of GM Aur. From
these observations, we extracted four measurements of the properties
of GM Aur.  First, we measured a lower accretion rate in GM\_3 than in GM\_1 and GM\_2.  For
GM\_3, we also measured a lower dust mass and a
lower \h2 density in the inner disk than the earlier GM\_1 and GM\_2 data. Lastly, the
change in the FUV--NIR emission of GM Aur was observed to occur over a timescale of 3 months.  In the the following, we consider which physical
properties of the system may have been different during GM\_3 to
explain our observations.  In particular, we focus on how
inhomogeneities in the inner disk may explain the four measured
properties of GM Aur.

We can explain the decrease in the NUV emission of GM Aur with our
accretion shock models, which measure a lower accretion rate in GM\_3.  It is important to define what a single measured accretion rate corresponds to.  Accretion rates are derived by measuring the 
shock emission in excess above the stellar photosphere
and using the accretion shock model (or a slab model)
to account for the emission outside the observed wavelengths 
(i.e., $<$1000 and $>$ 1 ${\micron}$ in this work).
Determining the accretion
luminosity from the excess emission, and using the stellar
mass and radius, the accretion rate then follows
\citep[e.g.,][]{gullbring98, herczeg08, ingleby13}. 
And so, the reported accretion rate is a measure of
the emission created from mass impacting 
a region of the stellar surface at a
particular point in time, which in turn corresponds to a particular
orientation of the star (and hence magnetosphere) and disk along our
line of sight.  
Analysis of the spectral energy distribution of the excess
emission in \S \ref{accretion}
indicates that in GM\_3 
we were observing a magnetospheric 
region characterized by low density 
accretion columns. This does not preclude
that high density columns were
present at other azimuthal regions that were not visible at that time.

Considering 5 epochs of existing STIS spectra (combining the current observations with those obtained in 2003 and 2010), 4 out of 5 times GM Aur was observed during a 
high state
of accretion and in GM\_3 we caught it in a low state. 
Compared to results from accretion shock
fitting of our multi-epoch data, the accretion rates in 2003 and 2010 were
similar to the higher accretion rates of GM\_1 and GM\_2 (see $\S$\ref{fuv}).  
Specifically, in the multi-column fitting of the spectra, all but  the GM\_3 epoch are characterized by high density accretion columns with small filling factors.
As discussed in \citet{ingleby13}, lower density columns
could still be contributing, but their emission could not
be extracted from the brighter stellar spectrum.  Whereas during GM\_3 the accretion columns had lower densities with larger stellar surface coverage.
Although schematic, this evidence points to an inhomogeneous and variable magnetosphere. This is not surprising
given the highly complex nature of the stellar magnetic fields determined by
Zeeman Doppler techniques 
\citep{donati08,donati11,donati13,gregory11}.
However, in addition
to the complexity of the magnetic fields, the structure
of the inner disk from which the mass is coming from must
play a role in the driving the full spectrum of variability.

Simultaneous with the NUV decrease in GM\_3, there are
decreases in the NIR spectrum and FUV \h2 emission feature at 1600
{\AA}.  We attribute these to lower dust masses and lower \h2 densities in the inner disk,
respectively.  In \S \ref{ir} we discussed the optically thin
dust models used to fit the NIR excess and found that the dust mass was
a factor of 1.4 lower in GM\_3 than in GM\_1.  Since the dust is
optically thin, this means the dust surface density decreased by a factor of 1.4.  Interestingly, 1.4 is the same factor by which we estimate the \h2
density decreased between GM\_1 and GM\_3 in
\S \ref{fuv}.  This suggests that the \h2 1600 {\AA} feature
 is tracing the same location in the disk as the
dust, and that the dust and gas are
well-mixed in the inner disk.  The decrease in the \h2
emission feature at 1600 {\AA} is not accompanied by a change in \h2
fluorescent lines, which should also trace the gas; however, this may
be due to different origins of
the \h2 lines.  As discussed earlier (\S \ref{fuv}), the \h2 emission feature at 1600
{\AA} is a consequence of collisional excitation following X-ray
ionization of heavy elements in the inner disk \citep{abgrall97,bergin04} while fluorescent lines are due to excitation of
electrons to high electronic levels of \h2 
by UV Ly$\alpha$ emission \citep{herczeg02}.  X-rays penetrate deeper into the disk than UV photons
because UV radiation is absorbed by small dust grains in
the uppermost layers of the disk \citep{nomura07}.  We expect that the \h2 emission feature at 1600 {\AA} is coming from
deeper in the disk and is tracing the same disk regions as seen
in the NIR dust emission.

Our analysis suggests a link between the mass in gas and dust in the disk and the mass being loaded onto the star at the magnetospheric truncation radius.  However, the kinematic timescales where the gas and dust reside in the disk are problematic given the 3 month time difference between GM\_1 and GM\_3.  At the dust destruction radius, the viscous timescale is $\sim$ 2--10 years using
Equation 37 of \citet{hartmann98} with R$_{1}$ = 10 R$_{*}$,
T=5000-1000, and $\alpha$=0.01.  It is unlikely that the mass in the
entire inner disk is decreasing within 3 months and that this is
causing the lower accretion rate in GM\_3 given that the dust and gas
in the inner disk are fed by the outer disk on the viscous timescale.  A period of 3 months requires a change in the disk to occur near the magnetospheric radius, $\sim$5 $R_{\ast}$.  At this radius, the viscous timescale is no longer relevant as material is transported inward along magnetic field lines as opposed to viscously through the disk.

In order for a similar decline of mass in the dust and \h2, the two must be spatially related, but dust must reside outside the dust destruction radius (typically $\sim$ 10 R$_{\ast}$) where the viscous timescale is much longer than 3 months.  However, if we allow large dust grains, 1 mm, and a sublimation temperature of 1600 K (as opposed to the commonly used 1400 K), dust grains can survive down to 0.07 AU, following \citet{dalessio05}, which is approximately the corotation radius.  \citet{stauffer15} found that, for a sample of young stars, IR variability indicates that inner disk walls may be present near the corotation radius.  In our simple model, we assume the magnetospheric truncation radius is 5$R_{\ast}$ \citep{calvet98}, or 0.04 AU.  In reality, the truncation radius depends on the strength of the magnetic field and accretion rate.  In addition, magnetospheres are complex, therefore the truncation radius varies and may extend to corotation at times \citep{donati11}.  If some of the NIR emission is produced in the large grains near the midplane (allowing for small grains remaining in the upper layers), then the gas and dust may be co-spatial near the corotation radius and also linked to an extended magnetosphere.  \citet{blinova15} showed that if the magnetospheric truncation radius ($r_m$) is near the corotation radius ($r_{cor}$), $r_{cor}\sim1.4\; r_m$, then accretion is chaotic and variable, whereas for $r_{cor}>1.4\; r_m$, accretion is stable.  Unstable accretion occurs through short lived ``tongues", contributing to inhomogeneous accretion.

While the NUV emission is related to variable accretion, FUV and NIR variability is rooted in the disk.  We cannot conclusively say what is leading to inhomogeneities in the inner disk with our low cadence data set.  One theory is that disk material builds up near the corotation radius before emptying quickly onto the star \citep{dangelo12}.  This would produce a period of lower density in the inner disk after the material is dumped onto the star.  While this phenomenon is typically used to describe objects undergoing large accretion bursts (like EXors), \citet{dangelo12} note that any object with a strong magnetic field and low accretion rate should undergo this affect at some level.  As discussed, the magnetosphere may truncate the disk closer to corotation than typically assumed, and gas and large dust grains may coexist near corotation, linking the decline in disk mass observed with the lower accretion rate.  Alternatively, it is possible that inner disk gas may have an emission component in the NIR.  \citet{tannirkulam08} required the presence of NIR emission inside the dust destruction radius of Herbig Ae stars and suggested optically thick gas as the source.  If the NIR does come from inner disk gas, then the inhomogeneities and magnetospheric truncation radius may reside inside corotation.

Another common cause of variability in CTTS stems from variable extinction by dust.  GM Aur becomes redder as it dims during the GM\_3 epoch and objects with that behavior may be undergoing such variable dust extinction, possibly from warps in the circumstellar disk \citep{gunther14,cody14}.  Variable obscuration requires that the disk be observed close to edge on, as in the well studied case of AA Tau \citep{bouvier13}.  GM Aur has a well determined inclination of $\sim$ 57$^{\circ}$ \citep{dutrey98}, therefore dust obscuration requires disk structures at large scale heights, unlikely to exist for any sustained period of time.  A dimming event in the CTTS RW Aur A (inclined between 45$^{\circ}$ and 60$^{\circ}$) may be due to dust carried to large scale heights above the disk in an enhanced wind, producing the obscuration \citep{petrov15}; however, RW Aur A is known to have strong outflows as evidenced by existing jets, whereas GM Aur shows no sign of jets.  In addition, \citet{ellerbroek14} showed that dust lifted to large scale heights above the disk, while producing dimming in the optical by obscuring the star, would increase the NIR emission as the surface area of dust is increased.

In conclusion, we propose that the observations presented here are
tracing inhomogeneities in the inner disk and magnetosphere.  Parsing out the detailed structure
of these regions and how they interact is an
extremely difficult task given the high complexity
\citep[e.g.,][]{romanova08, kulkarni08, romanova12, kurosawa13}.  In
fact, there are some observed properties of GM~Aur that we
cannot yet explain.  We note that while the dust mass and \h2 density
decrease by a factor of 1.4 in GM\_3, the accretion rate decreases by
a factor of 2.8 (although given the errors, these numbers may in fact be more similar).  The surface density may not scale exactly with $\mdot$ as factors such as temperature and viscosity also play a role.
It is puzzling what causes this difference, given
that it is not clear exactly how accretion occurs onto the star.  
In any event, our
data connect the disk, near the footprints of the magnetic field lines, to the material in the accretion shock at the stellar surface.  Higher cadence
observations of GM~Aur would be ideal to test our proposal and
to make further progress on the issue of variability of TTS.

\section{Summary}
We analyzed variability in the FUV to NIR spectrum of GM Aur over
3 epochs of observation, separated by one week and 3 months.  We found
that GM\_1 and GM\_2 were nearly identical in fluxes.
GM\_3 was observed during an epoch of low fluxes across the entire
spectrum.  We drew the following conclusions:
\begin{itemize}
\item
The accretion rate of GM\_3 was lower than the first two epochs by a
factor of 2.8.  While the surface coverage of the accretion columns
remained approximately constant, the energy flux in the columns
decreased in GM\_3.  Energy flux is related to the density, so GM\_3
had low densities in the accretion columns compared to the earlier
epochs.

\item
We fit SpeX observations, obtained contemporaneously with the STIS
spectra, with models of optically thin dust in the inner cavity of GM
Aur.  During the epoch of low accretion, the mass of dust in the disk
decreased by a factor of 1.4.

\item
The FUV spectrum indicates that the emission from 
\h2 molecules decreased
during the epoch of low accretion and decreased dust mass.  In
particular, we found that the spectrum of \h2 collisionally excited
by free electrons in the disk described the shape of the excess in
GM\_1 over GM\_3.  
Moreover, we find that the mass of \h2
changed by the same factor as the dust.

\item
GM~Aur is seen here to vary within a 3 month period, placing the origin of variability near the magnetospheric truncation radius.  If the disk is truncated near corotation and dust can survive near corotation in large grains, then inhomogeneities at that location in the disk can account for accretion variability.  The build up and subsequent dumping of material onto the star is one possible explanation for the source of disk inhomogeneities.

\end{itemize}

We conclude that these FUV-NIR data are providing a glimpse at the complexity
of the interaction between the stellar magnetic field and the inner disk. Future
observational work is needed to test this conclusion as well as theoretical work
on the observable properties of star-disk interactions.

\section{Acknowledgments}
LI and NC acknowledge support from 
HST grant No. GO-11608. The authors thank Hal Levinson, Andrew Youdin and Kevin Flaherty for insightful discussions.  LI would like to thank the anonymous referee for useful comments and suggestions.   The authors acknowledge Daryl Kim for his critical role in the BASS observing program.  This work is supported at The Aerospace Corporation by the Independent Research and Development Program.


\begin{thebibliography}{85}
\expandafter\ifx\csname natexlab\endcsname\relax\def\natexlab#1{#1}\fi

\bibitem[{{Abgrall} {et~al.}(1997){Abgrall}, {Roueff}, {Liu}, \&
  {Shemansky}}]{abgrall97}
{Abgrall}, H., {Roueff}, E., {Liu}, X., \& {Shemansky}, D.~E. 1997, \apj, 481,
  557

\bibitem[{{Alencar} {et~al.}(2010){Alencar}, {Teixeira}, {Guimar{\~a}es},
  {McGinnis}, {Gameiro}, {Bouvier}, {Aigrain}, {Flaccomio}, \&
  {Favata}}]{alencar10}
{Alencar}, S.~H.~P., {Teixeira}, P.~S., {Guimar{\~a}es}, M.~M., {et~al.} 2010,
  \aap, 519, A88

\bibitem[{{Andrews} {et~al.}(2011){Andrews}, {Wilner}, {Espaillat}, {Hughes},
  {Dullemond}, {McClure}, {Qi}, \& {Brown}}]{andrews11}
{Andrews}, S.~M., {Wilner}, D.~J., {Espaillat}, C., {et~al.} 2011, \apj, 732,
  42

\bibitem[{{Ardila} {et~al.}(2013){Ardila}, {Herczeg}, {Gregory}, {Ingleby},
  {France}, {Brown}, {Edwards}, {Johns-Krull}, {Linsky}, {Yang}, {Valenti},
  {Abgrall}, {Alexander}, {Bergin}, {Bethell}, {Brown}, {Calvet}, {Espaillat},
  {Hillenbrand}, {Hussain}, {Roueff}, {Schindhelm}, \& {Walter}}]{ardila13}
{Ardila}, D.~R., {Herczeg}, G.~J., {Gregory}, S.~G., {et~al.} 2013, \apjs, 207,
  1

\bibitem[{{Bergin} {et~al.}(2004){Bergin}, {Calvet}, {Sitko}, {Abgrall},
  {D'Alessio}, {Herczeg}, {Roueff}, {Qi}, {Lynch}, {Russell}, {Brafford}, \&
  {Perry}}]{bergin04}
{Bergin}, E., {Calvet}, N., {Sitko}, M.~L., {et~al.} 2004, \apjl, 614, L133

\bibitem[{{Blinova} {et~al.}(2015){Blinova}, {Romanova}, \&
  {Lovelace}}]{blinova15}
{Blinova}, A.~A., {Romanova}, M.~M., \& {Lovelace}, R.~V.~E. 2015, ArXiv
  e-prints

\bibitem[{{Bouvier} {et~al.}(2013){Bouvier}, {Grankin}, {Ellerbroek}, {Bouy},
  \& {Barrado}}]{bouvier13}
{Bouvier}, J., {Grankin}, K., {Ellerbroek}, L.~E., {Bouy}, H., \& {Barrado}, D.
  2013, \aap, 557, A77

\bibitem[{{Bouvier} {et~al.}(1999){Bouvier}, {Chelli}, {Allain}, {Carrasco},
  {Costero}, {Cruz-Gonzalez}, {Dougados}, {Fern{\'a}ndez}, {Mart{\'{\i}}n},
  {M{\'e}nard}, {Mennessier}, {Mujica}, {Recillas}, {Salas}, {Schmidt}, \&
  {Wichmann}}]{bouvier99}
{Bouvier}, J., {Chelli}, A., {Allain}, S., {et~al.} 1999, \aap, 349, 619

\bibitem[{{Calvet} {et~al.}(2002){Calvet}, {D'Alessio}, {Hartmann}, {Wilner},
  {Walsh}, \& {Sitko}}]{calvet02}
{Calvet}, N., {D'Alessio}, P., {Hartmann}, L., {et~al.} 2002, \apj, 568, 1008

\bibitem[{{Calvet} \& {Gullbring}(1998)}]{calvet98}
{Calvet}, N., \& {Gullbring}, E. 1998, \apj, 509, 802

\bibitem[{{Calvet} {et~al.}(2004){Calvet}, {Muzerolle}, {Brice{\~n}o},
  {Hern{\'a}ndez}, {Hartmann}, {Saucedo}, \& {Gordon}}]{calvet04}
{Calvet}, N., {Muzerolle}, J., {Brice{\~n}o}, C., {et~al.} 2004, \aj, 128, 1294

\bibitem[{{Calvet} {et~al.}(2005){Calvet}, {D'Alessio}, {Watson},
  {Franco-Hern{\'a}ndez}, {Furlan}, {Green}, {Sutter}, {Forrest}, {Hartmann},
  {Uchida}, {Keller}, {Sargent}, {Najita}, {Herter}, {Barry}, \&
  {Hall}}]{calvet05b}
{Calvet}, N., {D'Alessio}, P., {Watson}, D.~M., {et~al.} 2005, \apjl, 630, L185

\bibitem[{{Chen} {et~al.}(2006){Chen}, {Sargent}, {Bohac}, {Kim},
  {Leibensperger}, {Jura}, {Najita}, {Forrest}, {Watson}, {Sloan}, \&
  {Keller}}]{chen06}
{Chen}, C.~H., {Sargent}, B.~A., {Bohac}, C., {et~al.} 2006, \apjs, 166, 351

\bibitem[{{Chiang} \& {Goldreich}(1999)}]{chiang99}
{Chiang}, E.~I., \& {Goldreich}, P. 1999, \apj, 519, 279

\bibitem[{{Chou} {et~al.}(2013){Chou}, {Takami}, {Manset}, {Beck}, {Pyo},
  {Chen}, {Panwar}, {Karr}, {Shang}, \& {Liu}}]{chou13}
{Chou}, M.-Y., {Takami}, M., {Manset}, N., {et~al.} 2013, \aj, 145, 108

\bibitem[{{Cody} {et~al.}(2014){Cody}, {Stauffer}, {Baglin}, {Micela},
  {Rebull}, {Flaccomio}, {Morales-Calder{\'o}n}, {Aigrain}, {Bouvier},
  {Hillenbrand}, {Gutermuth}, {Song}, {Turner}, {Alencar}, {Zwintz},
  {Plavchan}, {Carpenter}, {Findeisen}, {Carey}, {Terebey}, {Hartmann},
  {Calvet}, {Teixeira}, {Vrba}, {Wolk}, {Covey}, {Poppenhaeger}, {G{\"u}nther},
  {Forbrich}, {Whitney}, {Affer}, {Herbst}, {Hora}, {Barrado}, {Holtzman},
  {Marchis}, {Wood}, {Medeiros Guimar{\~a}es}, {Lillo Box}, {Gillen},
  {McQuillan}, {Espaillat}, {Allen}, {D'Alessio}, \& {Favata}}]{cody14}
{Cody}, A.~M., {Stauffer}, J., {Baglin}, A., {et~al.} 2014, \aj, 147, 82

\bibitem[{{Costigan} {et~al.}(2012){Costigan}, {Scholz}, {Stelzer}, {Ray},
  {Vink}, \& {Mohanty}}]{costigan12}
{Costigan}, G., {Scholz}, A., {Stelzer}, B., {et~al.} 2012, ArXiv e-prints

\bibitem[{{Cushing} {et~al.}(2004){Cushing}, {Vacca}, \& {Rayner}}]{cushing04}
{Cushing}, M.~C., {Vacca}, W.~D., \& {Rayner}, J.~T. 2004, \pasp, 116, 362

\bibitem[{{D'Alessio} {et~al.}(2005){D'Alessio}, {Hartmann}, {Calvet},
  {Franco-Hern{\'a}ndez}, {Forrest}, {Sargent}, {Furlan}, {Uchida}, {Green},
  {Watson}, {Chen}, {Kemper}, {Sloan}, \& {Najita}}]{dalessio05}
{D'Alessio}, P., {Hartmann}, L., {Calvet}, N., {et~al.} 2005, \apj, 621, 461

\bibitem[{{D'Angelo} \& {Spruit}(2012)}]{dangelo12}
{D'Angelo}, C.~R., \& {Spruit}, H.~C. 2012, \mnras, 420, 416

\bibitem[{{Donati} {et~al.}(2008){Donati}, {Jardine}, {Gregory}, {Petit},
  {Paletou}, {Bouvier}, {Dougados}, {M{\'e}nard}, {Collier Cameron}, {Harries},
  {Hussain}, {Unruh}, {Morin}, {Marsden}, {Manset}, {Auri{\`e}re}, {Catala}, \&
  {Alecian}}]{donati08}
{Donati}, J.-F., {Jardine}, M.~M., {Gregory}, S.~G., {et~al.} 2008, \mnras,
  386, 1234

\bibitem[{{Donati} {et~al.}(2011){Donati}, {Bouvier}, {Walter}, {Gregory},
  {Skelly}, {Hussain}, {Flaccomio}, {Argiroffi}, {Grankin}, {Jardine},
  {M{\'e}nard}, {Dougados}, \& {Romanova}}]{donati11}
{Donati}, J.-F., {Bouvier}, J., {Walter}, F.~M., {et~al.} 2011, \mnras, 412,
  2454

\bibitem[{{Donati} {et~al.}(2013){Donati}, {Gregory}, {Alencar}, {Hussain},
  {Bouvier}, {Jardine}, {M{\'e}nard}, {Dougados}, {Romanova}, \& {MaPP
  Collaboration}}]{donati13}
{Donati}, J.-F., {Gregory}, S.~G., {Alencar}, S.~H.~P., {et~al.} 2013, \mnras,
  436, 881

\bibitem[{{Dupree} {et~al.}(2014){Dupree}, {Brickhouse}, {Cranmer}, {Berlind},
  {Strader}, \& {Smith}}]{dupree14}
{Dupree}, A.~K., {Brickhouse}, N.~S., {Cranmer}, S.~R., {et~al.} 2014, \apj,
  789, 27

\bibitem[{{Dutrey} {et~al.}(1998){Dutrey}, {Guilloteau}, {Prato}, {Simon},
  {Duvert}, {Schuster}, \& {Menard}}]{dutrey98}
{Dutrey}, A., {Guilloteau}, S., {Prato}, L., {et~al.} 1998, \aap, 338, L63

\bibitem[{{Edwards} {et~al.}(2006){Edwards}, {Fischer}, {Hillenbrand}, \&
  {Kwan}}]{edwards06}
{Edwards}, S., {Fischer}, W., {Hillenbrand}, L., \& {Kwan}, J. 2006, \apj, 646,
  319

\bibitem[{{Ellerbroek} {et~al.}(2014){Ellerbroek}, {Podio}, {Dougados},
  {Cabrit}, {Sitko}, {Sana}, {Kaper}, {de Koter}, {Klaassen}, {Mulders},
  {Mendigut{\'{\i}}a}, {Grady}, {Grankin}, {van Winckel}, {Bacciotti},
  {Russell}, {Lynch}, {Hammel}, {Beerman}, {Day}, {Huelsman}, {Werren},
  {Henden}, \& {Grindlay}}]{ellerbroek14}
{Ellerbroek}, L.~E., {Podio}, L., {Dougados}, C., {et~al.} 2014, \aap, 563, A87

\bibitem[{{Espaillat} {et~al.}(2007){Espaillat}, {Calvet}, {D'Alessio},
  {Hern{\'a}ndez}, {Qi}, {Hartmann}, {Furlan}, \& {Watson}}]{espaillat07}
{Espaillat}, C., {Calvet}, N., {D'Alessio}, P., {et~al.} 2007, \apjl, 670, L135

\bibitem[{{Espaillat} {et~al.}(2011){Espaillat}, {Furlan}, {D'Alessio},
  {Sargent}, {Nagel}, {Calvet}, {Watson}, \& {Muzerolle}}]{espaillat11}
{Espaillat}, C., {Furlan}, E., {D'Alessio}, P., {et~al.} 2011, \apj, 728, 49

\bibitem[{{Espaillat} {et~al.}(2010){Espaillat}, {D'Alessio}, {Hern{\'a}ndez},
  {Nagel}, {Luhman}, {Watson}, {Calvet}, {Muzerolle}, \&
  {McClure}}]{espaillat10}
{Espaillat}, C., {D'Alessio}, P., {Hern{\'a}ndez}, J., {et~al.} 2010, \apj,
  717, 441

\bibitem[{{Espaillat} {et~al.}(2014){Espaillat}, {Muzerolle}, {Najita},
  {Andrews}, {Zhu}, {Calvet}, {Kraus}, {Hashimoto}, {Kraus}, \&
  {D'Alessio}}]{espaillat14}
{Espaillat}, C., {Muzerolle}, J., {Najita}, J., {et~al.} 2014, ArXiv e-prints

\bibitem[{{Fischer} {et~al.}(2011){Fischer}, {Edwards}, {Hillenbrand}, \&
  {Kwan}}]{fischer11}
{Fischer}, W., {Edwards}, S., {Hillenbrand}, L., \& {Kwan}, J. 2011, \apj, 730,
  73

\bibitem[{{France} {et~al.}(2011{\natexlab{a}}){France}, {Yang}, \&
  {Linsky}}]{france11a}
{France}, K., {Yang}, H., \& {Linsky}, J.~L. 2011{\natexlab{a}}, \apj, 729, 7

\bibitem[{{France} {et~al.}(2011{\natexlab{b}}){France}, {Schindhelm}, {Burgh},
  {Herczeg}, {Harper}, {Brown}, {Green}, {Linsky}, {Yang}, {Abgrall}, {Ardila},
  {Bergin}, {Bethell}, {Brown}, {Calvet}, {Espaillat}, {Gregory},
  {Hillenbrand}, {Hussain}, {Ingleby}, {Johns-Krull}, {Roueff}, {Valenti}, \&
  {Walter}}]{france11b}
{France}, K., {Schindhelm}, E., {Burgh}, E.~B., {et~al.} 2011{\natexlab{b}},
  \apj, 734, 31

\bibitem[{{Gregory} \& {Donati}(2011)}]{gregory11}
{Gregory}, S.~G., \& {Donati}, J.-F. 2011, Astronomische Nachrichten, 332, 1027

\bibitem[{{Gullbring} {et~al.}(1998){Gullbring}, {Hartmann}, {Briceno}, \&
  {Calvet}}]{gullbring98}
{Gullbring}, E., {Hartmann}, L., {Briceno}, C., \& {Calvet}, N. 1998, \apj,
  492, 323

\bibitem[{{G{\"u}nther} {et~al.}(2014){G{\"u}nther}, {Cody}, {Covey},
  {Hillenbrand}, {Plavchan}, {Poppenhaeger}, {Rebull}, {Stauffer}, {Wolk},
  {Allen}, {Bayo}, {Gutermuth}, {Hora}, {Meng}, {Morales-Calder{\'o}n},
  {Parks}, \& {Song}}]{gunther14}
{G{\"u}nther}, H.~M., {Cody}, A.~M., {Covey}, K.~R., {et~al.} 2014, \aj, 148,
  122

\bibitem[{{Hartigan} {et~al.}(1991){Hartigan}, {Kenyon}, {Hartmann}, {Strom},
  {Edwards}, {Welty}, \& {Stauffer}}]{hartigan91}
{Hartigan}, P., {Kenyon}, S.~J., {Hartmann}, L., {et~al.} 1991, \apj, 382, 617

\bibitem[{{Hartmann} {et~al.}(1998){Hartmann}, {Calvet}, {Gullbring}, \&
  {D'Alessio}}]{hartmann98}
{Hartmann}, L., {Calvet}, N., {Gullbring}, E., \& {D'Alessio}, P. 1998, \apj,
  495, 385

\bibitem[{{Herbst} {et~al.}(1994){Herbst}, {Herbst}, {Grossman}, \&
  {Weinstein}}]{herbst94}
{Herbst}, W., {Herbst}, D.~K., {Grossman}, E.~J., \& {Weinstein}, D. 1994, \aj,
  108, 1906

\bibitem[{{Herczeg} \& {Hillenbrand}(2008)}]{herczeg08}
{Herczeg}, G.~J., \& {Hillenbrand}, L.~A. 2008, \apj, 681, 594

\bibitem[{{Herczeg} \& {Hillenbrand}(2014)}]{herczeg14}
---. 2014, \apj, 786, 97

\bibitem[{{Herczeg} {et~al.}(2002){Herczeg}, {Linsky}, {Valenti},
  {Johns-Krull}, \& {Wood}}]{herczeg02}
{Herczeg}, G.~J., {Linsky}, J.~L., {Valenti}, J.~A., {Johns-Krull}, C.~M., \&
  {Wood}, B.~E. 2002, \apj, 572, 310

\bibitem[{{Herczeg} {et~al.}(2004){Herczeg}, {Wood}, {Linsky}, {Valenti}, \&
  {Johns-Krull}}]{herczeg04}
{Herczeg}, G.~J., {Wood}, B.~E., {Linsky}, J.~L., {Valenti}, J.~A., \&
  {Johns-Krull}, C.~M. 2004, \apj, 607, 369

\bibitem[{{Hughes} {et~al.}(2009){Hughes}, {Andrews}, {Espaillat}, {Wilner},
  {Calvet}, {D'Alessio}, {Qi}, {Williams}, \& {Hogerheijde}}]{hughes09}
{Hughes}, A.~M., {Andrews}, S.~M., {Espaillat}, C., {et~al.} 2009, \apj, 698,
  131

\bibitem[{{Ingleby} {et~al.}(2012){Ingleby}, {Calvet}, {Herczeg}, \&
  {Brice{\~n}o}}]{ingleby12}
{Ingleby}, L., {Calvet}, N., {Herczeg}, G., \& {Brice{\~n}o}, C. 2012, \apjl,
  752, L20

\bibitem[{{Ingleby} {et~al.}(2009){Ingleby}, {Calvet}, {Bergin}, {Yerasi},
  {Espaillat}, {Herczeg}, {Roueff}, {Abgrall}, {Hern{\'a}ndez}, {Brice{\~n}o},
  {Pascucci}, {Miller}, {Fogel}, {Hartmann}, {Meyer}, {Carpenter}, {Crockett},
  \& {McClure}}]{ingleby09}
{Ingleby}, L., {Calvet}, N., {Bergin}, E., {et~al.} 2009, \apjl, 703, L137

\bibitem[{{Ingleby} {et~al.}(2011){Ingleby}, {Calvet}, {Bergin}, {Herczeg},
  {Brown}, {Alexander}, {Edwards}, {Espaillat}, {France}, {Gregory},
  {Hillenbrand}, {Roueff}, {Valenti}, {Walter}, {Johns-Krull}, {Brown},
  {Linsky}, {McClure}, {Ardila}, {Abgrall}, {Bethell}, {Hussain}, \&
  {Yang}}]{ingleby11b}
---. 2011, \apj, 743, 105

\bibitem[{{Ingleby} {et~al.}(2013){Ingleby}, {Calvet}, {Herczeg}, {Blaty},
  {Walter}, {Ardila}, {Alexander}, {Edwards}, {Espaillat}, {Gregory},
  {Hillenbrand}, \& {Brown}}]{ingleby13}
{Ingleby}, L., {Calvet}, N., {Herczeg}, G., {et~al.} 2013, \apj, 767, 112

\bibitem[{{Johns-Krull} {et~al.}(2000){Johns-Krull}, {Valenti}, \&
  {Linsky}}]{johnskrull00}
{Johns-Krull}, C.~M., {Valenti}, J.~A., \& {Linsky}, J.~L. 2000, \apj, 539, 815

\bibitem[{{Kenyon} \& {Hartmann}(1995)}]{kenyon95}
{Kenyon}, S.~J., \& {Hartmann}, L. 1995, \apjs, 101, 117

\bibitem[{{Koenigl}(1991)}]{koenigl91}
{Koenigl}, A. 1991, \apjl, 370, L39

\bibitem[{{Koerner} {et~al.}(1993){Koerner}, {Sargent}, \&
  {Beckwith}}]{koerner93}
{Koerner}, D.~W., {Sargent}, A.~I., \& {Beckwith}, S.~V.~W. 1993, \icarus, 106,
  2

\bibitem[{{Kulkarni} \& {Romanova}(2008)}]{kulkarni08}
{Kulkarni}, A.~K., \& {Romanova}, M.~M. 2008, \mnras, 386, 673

\bibitem[{{Kurosawa} \& {Romanova}(2013)}]{kurosawa13}
{Kurosawa}, R., \& {Romanova}, M.~M. 2013, \mnras, 431, 2673

\bibitem[{{Long} {et~al.}(2008){Long}, {Romanova}, \& {Lovelace}}]{long08}
{Long}, M., {Romanova}, M.~M., \& {Lovelace}, R.~V.~E. 2008, \mnras, 386, 1274

\bibitem[{{Manara} {et~al.}(2014){Manara}, {Testi}, {Natta}, {Rosotti},
  {Benisty}, {Ercolano}, \& {Ricci}}]{manara14}
{Manara}, C.~F., {Testi}, L., {Natta}, A., {et~al.} 2014, ArXiv e-prints

\bibitem[{{Marsh} \& {Mahoney}(1992)}]{marsh92}
{Marsh}, K.~A., \& {Mahoney}, M.~J. 1992, \apjl, 395, L115

\bibitem[{{Matthews} {et~al.}(2014){Matthews}, {Krivov}, {Wyatt}, {Bryden}, \&
  {Eiroa}}]{matthews14}
{Matthews}, B.~C., {Krivov}, A.~V., {Wyatt}, M.~C., {Bryden}, G., \& {Eiroa},
  C. 2014, Protostars and Planets VI

\bibitem[{{McClure} {et~al.}(2013){McClure}, {Calvet}, {Espaillat}, {Hartmann},
  {Hern{\'a}ndez}, {Ingleby}, {Luhman}, {D'Alessio}, \& {Sargent}}]{mcclure13a}
{McClure}, M.~K., {Calvet}, N., {Espaillat}, C., {et~al.} 2013, \apj, 769, 73

\bibitem[{{Micela} \& {Marino}(2003)}]{micela03}
{Micela}, G., \& {Marino}, A. 2003, \aap, 404, 637

\bibitem[{{Morales-Calder{\'o}n} {et~al.}(2011){Morales-Calder{\'o}n},
  {Stauffer}, {Hillenbrand}, {Gutermuth}, {Song}, {Rebull}, {Plavchan},
  {Carpenter}, {Whitney}, {Covey}, {Alves de Oliveira}, {Winston},
  {McCaughrean}, {Bouvier}, {Guieu}, {Vrba}, {Holtzman}, {Marchis}, {Hora},
  {Wasserman}, {Terebey}, {Megeath}, {Guinan}, {Forbrich}, {Hu{\'e}lamo},
  {Riviere-Marichalar}, {Barrado}, {Stapelfeldt}, {Hern{\'a}ndez}, {Allen},
  {Ardila}, {Bayo}, {Favata}, {James}, {Werner}, \& {Wood}}]{morales11}
{Morales-Calder{\'o}n}, M., {Stauffer}, J.~R., {Hillenbrand}, L.~A., {et~al.}
  2011, \apj, 733, 50

\bibitem[{{Nomura} {et~al.}(2007){Nomura}, {Aikawa}, {Tsujimoto}, {Nakagawa},
  \& {Millar}}]{nomura07}
{Nomura}, H., {Aikawa}, Y., {Tsujimoto}, M., {Nakagawa}, Y., \& {Millar}, T.~J.
  2007, \apj, 661, 334

\bibitem[{{Owen} {et~al.}(2012){Owen}, {Clarke}, \& {Ercolano}}]{owen12}
{Owen}, J.~E., {Clarke}, C.~J., \& {Ercolano}, B. 2012, \mnras, 422, 1880

\bibitem[{{Owen} {et~al.}(2010){Owen}, {Ercolano}, {Clarke}, \&
  {Alexander}}]{owen10}
{Owen}, J.~E., {Ercolano}, B., {Clarke}, C.~J., \& {Alexander}, R.~D. 2010,
  \mnras, 401, 1415

\bibitem[{{Petrov} {et~al.}(2015){Petrov}, {Gahm}, {Djupvik}, {Babina},
  {Artemenko}, \& {Grankin}}]{petrov15}
{Petrov}, P.~P., {Gahm}, G.~F., {Djupvik}, A.~A., {et~al.} 2015, ArXiv e-prints

\bibitem[{{Rayner} {et~al.}(2003){Rayner}, {Toomey}, {Onaka}, {Denault},
  {Stahlberger}, {Vacca}, {Cushing}, \& {Wang}}]{rayner03}
{Rayner}, J.~T., {Toomey}, D.~W., {Onaka}, P.~M., {et~al.} 2003, \pasp, 115,
  362

\bibitem[{{Rice} {et~al.}(2003){Rice}, {Wood}, {Armitage}, {Whitney}, \&
  {Bjorkman}}]{rice03}
{Rice}, W.~K.~M., {Wood}, K., {Armitage}, P.~J., {Whitney}, B.~A., \&
  {Bjorkman}, J.~E. 2003, \mnras, 342, 79

\bibitem[{{Rigliaco} {et~al.}(2011){Rigliaco}, {Natta}, {Randich}, {Testi}, \&
  {Biazzo}}]{rigliaco11}
{Rigliaco}, E., {Natta}, A., {Randich}, S., {Testi}, L., \& {Biazzo}, K. 2011,
  \aap, 525, A47

\bibitem[{{Romanova} {et~al.}(2008){Romanova}, {Kulkarni}, \&
  {Lovelace}}]{romanova08}
{Romanova}, M.~M., {Kulkarni}, A.~K., \& {Lovelace}, R.~V.~E. 2008, \apjl, 673,
  L171

\bibitem[{{Romanova} {et~al.}(2012){Romanova}, {Ustyugova}, {Koldoba}, \&
  {Lovelace}}]{romanova12}
{Romanova}, M.~M., {Ustyugova}, G.~V., {Koldoba}, A.~V., \& {Lovelace},
  R.~V.~E. 2012, \mnras, 421, 63

\bibitem[{{Schindhelm} {et~al.}(2012){Schindhelm}, {France}, {Burgh},
  {Herczeg}, {Green}, {Brown}, {Brown}, \& {Valenti}}]{schindhelm12}
{Schindhelm}, E., {France}, K., {Burgh}, E.~B., {et~al.} 2012, \apj, 746, 97

\bibitem[{{Shu} {et~al.}(1994){Shu}, {Najita}, {Ostriker}, {Wilkin}, {Ruden},
  \& {Lizano}}]{shu94}
{Shu}, F., {Najita}, J., {Ostriker}, E., {et~al.} 1994, \apj, 429, 781

\bibitem[{{Siess} {et~al.}(2000){Siess}, {Dufour}, \& {Forestini}}]{siess00}
{Siess}, L., {Dufour}, E., \& {Forestini}, M. 2000, \aap, 358, 593

\bibitem[{{Skrutskie} {et~al.}(1990){Skrutskie}, {Dutkevitch}, {Strom},
  {Edwards}, {Strom}, \& {Shure}}]{skrutskie90}
{Skrutskie}, M.~F., {Dutkevitch}, D., {Strom}, S.~E., {et~al.} 1990, \aj, 99,
  1187

\bibitem[{{Stauffer} {et~al.}(2014){Stauffer}, {Cody}, {Baglin}, {Alencar},
  {Rebull}, {Hillenbrand}, {Venuti}, {Turner}, {Carpenter}, {Plavchan},
  {Findeisen}, {Carey}, {Terebey}, {Morales-Calder{\'o}n}, {Bouvier}, {Micela},
  {Flaccomio}, {Song}, {Gutermuth}, {Hartmann}, {Calvet}, {Whitney}, {Barrado},
  {Vrba}, {Covey}, {Herbst}, {Furesz}, {Aigrain}, \& {Favata}}]{stauffer14}
{Stauffer}, J., {Cody}, A.~M., {Baglin}, A., {et~al.} 2014, \aj, 147, 83

\bibitem[{{Stauffer} {et~al.}(2015){Stauffer}, {Cody}, {McGinnis}, {Rebull},
  {Hillenbrand}, {Turner}, {Carpenter}, {Plavchan}, {Carey}, {Terebey},
  {Morales-Calder{\'o}n}, {Alencar}, {Bouvier}, {Venuti}, {Hartmann}, {Calvet},
  {Micela}, {Flaccomio}, {Song}, {Gutermuth}, {Barrado}, {Vrba}, {Covey},
  {Padgett}, {Herbst}, {Gillen}, {Lyra}, {Medeiros Guimaraes}, {Bouy}, \&
  {Favata}}]{stauffer15}
{Stauffer}, J., {Cody}, A.~M., {McGinnis}, P., {et~al.} 2015, ArXiv e-prints

\bibitem[{{Strom} {et~al.}(1989){Strom}, {Strom}, {Edwards}, {Cabrit}, \&
  {Skrutskie}}]{strom89}
{Strom}, K.~M., {Strom}, S.~E., {Edwards}, S., {Cabrit}, S., \& {Skrutskie},
  M.~F. 1989, \aj, 97, 1451

\bibitem[{{Tannirkulam} {et~al.}(2008){Tannirkulam}, {Monnier}, {Millan-Gabet},
  {Harries}, {Pedretti}, {ten Brummelaar}, {McAlister}, {Turner}, {Sturmann},
  \& {Sturmann}}]{tannirkulam08}
{Tannirkulam}, A., {Monnier}, J.~D., {Millan-Gabet}, R., {et~al.} 2008, \apjl,
  677, L51

\bibitem[{{Uchida} \& {Shibata}(1985)}]{uchida85}
{Uchida}, Y., \& {Shibata}, K. 1985, \pasj, 37, 515

\bibitem[{{Vacca} {et~al.}(2003){Vacca}, {Cushing}, \& {Rayner}}]{vacca03}
{Vacca}, W.~D., {Cushing}, M.~C., \& {Rayner}, J.~T. 2003, \pasp, 115, 389

\bibitem[{{Whittet} {et~al.}(2004){Whittet}, {Shenoy}, {Clayton}, \&
  {Gordon}}]{whittet04}
{Whittet}, D.~C.~B., {Shenoy}, S.~S., {Clayton}, G.~C., \& {Gordon}, K.~D.
  2004, \apj, 602, 291

\bibitem[{{Wolk} {et~al.}(2005){Wolk}, {Harnden}, {Flaccomio}, {Micela},
  {Favata}, {Shang}, \& {Feigelson}}]{wolk05}
{Wolk}, S.~J., {Harnden}, Jr., F.~R., {Flaccomio}, E., {et~al.} 2005, \apjs,
  160, 423

\bibitem[{{Zhu} {et~al.}(2011){Zhu}, {Nelson}, {Hartmann}, {Espaillat}, \&
  {Calvet}}]{zhu11}
{Zhu}, Z., {Nelson}, R.~P., {Hartmann}, L., {Espaillat}, C., \& {Calvet}, N.
  2011, \apj, 729, 47

\bibitem[{{Zuckerman}(2001)}]{zuckerman01}
{Zuckerman}, B. 2001, \araa, 39, 549

\end{thebibliography}

\begin{deluxetable}{lcc}
\tablewidth{0pt}
\tablecaption{Log of Observations
\label{tabobs}}
\tablehead{\colhead{Epoch}&\colhead{Telescope/ Instrument}& \colhead{Date of Obs}}
\startdata
GM\_1 & \emph{HST}/STIS& 09-11-2011\\
GM\_1 & IRTF/SpeX		  &09-11-2011\\
GM\_2					      & \emph{HST}/STIS& 09-17-2011\\
GM\_2 & IRTF/SpeX	  &09-18-2011\\
GM\_3				      & \emph{HST}/STIS& 01-05-2012\\
GM\_3& IRTF/SpeX	  &01-06-2012\\
\enddata
\end{deluxetable}


\begin{deluxetable}{lcccccccc}
\tablewidth{0pt}
\tablecaption{Results from Multi-Component Model Fits to UV and Optical Spectra
\label{tabacc}}
\tablehead{
\colhead{Object} &\colhead{$f(10^{10})$}&\colhead{$f(3\times10^{10})$}&\colhead{$f(10^{11})$}&\colhead{$f(3\times10^{11})$}&\colhead{$f(10^{12})$}&\colhead{$f_{total}$}&\colhead{Weighted $\curf$}&\colhead{$\mdot\; (\msunyr)$}}
\startdata
GM\_1&0&0&0.03&0&0.001&0.031&$1.3\times10^{11}$&$1.1\times10^{-8}$\\
GM\_2&0&0&0.02&0&0.001&0.021&$1.4\times10^{11}$&$8.5\times10^{-9}$\\
GM\_3&0&0.04&0&0.0006&0&0.041&$3.4\times10^{10}$&$3.9\times10^{-9}$\\
\enddata
\tablecomments{The values in parentheses represents the energy flux ($\curf$) of each column in erg s$^{-1}$ cm$^{-3}$.}
\end{deluxetable}


\end{document}